\documentclass[aps,prl,twocolumn,floatfix,amssymb,amsmath,superscriptaddress,preprintnumbers]{revtex4-2}

\usepackage{graphicx}
\usepackage{tikz}
\usepackage{physics}
\usepackage{siunitx}
\begin{document}

\title{Matter Wave Analog of a Fiber-Optic Gyroscope}
\preprint{LA-UR-22-20582}

\author{Katarzyna A. Krzyzanowska}
\affiliation{Materials Physics and Applications Division, Los Alamos National Laboratory, Los Alamos, NM 87545, USA}
\author{Jorge Ferreras}
\affiliation{Materials Physics and Applications Division, Los Alamos National Laboratory, Los Alamos, NM 87545, USA}
\affiliation{Rutherford Appleton Laboratory, Didcot OX11 0QX, UK}
\author{Changhyun Ryu}
\affiliation{Materials Physics and Applications Division, Los Alamos National Laboratory, Los Alamos, NM 87545, USA}
\author{Edward Carlo Samson}
\affiliation{Materials Physics and Applications Division, Los Alamos National Laboratory, Los Alamos, NM 87545, USA}
\affiliation{Department of Physics, Miami University, Oxford, OH 45056, USA}
\author{Malcolm G. Boshier}
\affiliation{Materials Physics and Applications Division, Los Alamos National Laboratory, Los Alamos, NM 87545, USA}

\date{\today}

\begin{abstract}
Confining the propagating wavepackets of an atom interferometer inside a waveguide can substantially reduce the size of the device while preserving high sensitivity. We have realized a two-dimensional Sagnac atom interferometer in which Bose-condensed $^{87}$Rb atoms propagate within a tight waveguide formed by a collimated laser beam, a matter wave analog of the fiber optic gyro (FOG). The condensate is split, reflected, and recombined with a series of Bragg pulses while the waveguide moves transversely so that the wavepacket trajectories enclose an area. Delta-kick cooling is used to prepare low-density atomic wavepackets with a temperature of \SI{3}{\nano \kelvin}. The low density reduces the impact of interatomic interactions, while the low temperature limits the expansion of the wavepacket during the interferometer cycle. The effective enclosed area is \SI{0.8}{\milli \metre \squared} with an average fringe contrast of 20\% and underlying contrast up to 60\%. The main source of the reduced average contrast is phase noise caused by mechanical vibrations of the optical components. We present the first measurement of Allan deviation for such an atom rotation sensor, showing that the interferometer phase noise falls with averaging time $\tau$ as $\tau^{-1/2}$ for $\tau$ up to 10,000 seconds. The statistical noise falls below the Earth rotation rate after 30 minutes of averaging.
\end{abstract}

\maketitle

\section{Introduction}
Atom interferometers \cite{CarnalPRL1991, KeithPRL1991} have important applications in fundamental science \cite{bongs2019taking, CroninRMP2009} and as accelerometers, gravimeters, and rotation sensors \cite{CroninRMP2009, barrett2014the}. The latter rely on the Sagnac effect: rotation at rate $\vb{\Omega}$ of an atom interferometer whose arms enclose area $A$ creates a phase shift $\varphi_{Sg} = \frac{2m}{\hbar} \vb{\Omega} \cdot \vb{A}$, where $m$ is the atomic mass \cite{RiehlePRL1991}. While free-space atom interferometers have demonstrated the high sensitivity of this technique \cite{DurfeePRL2006}, their size must grow quadratically with interrogation time to accommodate free-fall under gravity. This ultimate limitation has motivated numerous efforts to confine propagating matter waves in tight waveguides \cite{dumke2002interferometer,japha2007using,muller2008towards,hui2012guided, RyuNJoP2015,navez2016matter,alzar2019compact,moukouri2021multipass}, analogous to light in the fiber optic gyro (FOG) \cite{vali1976fiber}, to deliver long interrogation times and large enclosed total area, and hence large sensitivity, in a small physics package. These factors are especially important for inertial navigation and other applications that require a compact transportable rotation sensor.

An atom interferometer gyro's performance limit is determined by its rotation sensitivity (proportional to the Sagnac area $A$) and by the quantum projection noise set by the particle flux and measurement rate \cite{ItanoPRA1993}.
The Angular Random Walk (ARW) of a shot noise-limited atom interferometer is $ARW = \hbar/2 m A \sqrt{N} \sqrt{\nu}$, where $N$ is the number of atoms and $\nu$ is the measurement rate. With a circular trajectory for atoms moving with velocity $v$ and a repetition rate limited by an interrogation time $T$, this expression becomes $ARW = \hbar \pi/(2 m v^2 T^{3/2} \sqrt{N})$.
Typically, atom interferometers can achieve a signal to noise ratio (SNR) of the order of \num{e2}, corresponding to the shot noise limit for $N=\num{e4}$ atoms \cite{TemplierSciAdv2022, NarducciAdvInPhys2022}. Improving the SNR further will require better atom detection techniques and precise noise control, which are immensely difficult for the current state of the art. For $N=\SI{e4}{atoms}$, a shot noise-limited $^{87}$Rb atom interferometer would need an interrogation time of $T=\SI{1.5}{\second}$ to achieve the ARW of \SI{e-4}{\degree \per \sqrt{\hour}} demonstrated by the best FOGs \cite{GuattariInertialSensSys2016, KorkishkoIEEEsymposium2018}. Much longer interrogation times of several seconds are of course desirable to deliver even better performance. While achieving such interrogation times for thermal atoms is challenging because of the unavoidable wavepacket dispersion, the narrow momentum width of BECs allows for extended interaction times and enables Bragg diffraction into a single momentum order \cite{WuPRA2005, WangPRL2005, GarciaPRA2006}. Therefore, a tight guide and BEC wavepackets are essential components of a practical matter wave analog of a FOG.

The first approach towards this goal of realizing the matter wave analog of a FOG used cold thermal atoms in guiding potentials formed by weak (few Hz) magnetic traps \cite{WuPRL2007, LuQiAPL2017}. In \cite{WuPRL2007}, short Bragg pulses created a thin phase grating which diffracted atoms into multiple momentum orders. The resulting trajectories formed closed loops when the guide was translated during the interferometer cycle, resulting in an atomic density grating pattern which could be probed by Bragg scattering. The pairing of BEC and Bragg diffraction was used in \cite{BurkePRA2009} to realize an atom interferometer enclosing \SI{0.1}{\milli \metre \squared} by imparting transverse motion to the moving wavepackets through a sudden displacement of the relatively weak (\SI{6}{\hertz}) harmonic trap.   Other approaches to creating trapped Sagnac atom interferometers where the splitting is based on manipulations of the potential have demonstrated much smaller enclosed area \cite{JoPRL2007} and \cite{RyuNJoP2015}. Recently, a novel architecture where BEC wavepackets are displaced to opposite sides of a weak harmonic potential and subsequently launched into circular trajectories has been shown to achieve \SI{0.5}{\milli \metre \squared} enclosed area \cite{MoanPRL2020}.

In this paper we report the first tightly-guided BEC Sagnac atom interferometer. The tight waveguide confinement perpendicular to the wavepacket velocity makes it analogous to the FOG and will make the device much less sensitive to platform motion than the weakly-trapped approaches discussed above. The interferometer wavepackets are trapped by the optical dipole potential of a collimated laser beam, enabling use of an $m_F=0$ state which is first-order insensitive to magnetic fields and hence to magnetic field gradients. Also, with the optical waveguide, phase fluctuations due to guiding potential fluctuations are mostly common-mode. These characteristics make the optically-guided interferometer much more robust against environment magnetic field noise than its magnetically-guided counterpart. The new design also has a straightforward extension to multi-axis measurement without requiring that the sensor be rotated or that gravity be compensated. In addition to reporting the biggest Sagnac area reported to date, \SI{0.8}{\milli \metre \squared }, we also present the first Allan deviation measurement for a guided Sagnac atom interferometer along with an in-depth study of the noise behavior of the device.

\begin{figure}
\begin{tikzpicture}
\node[inner sep=0pt] at (0,1.2) {     \includegraphics[width=0.48\textwidth]{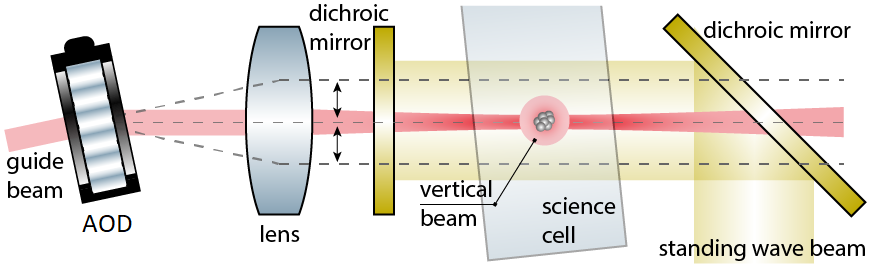}     };
\node[inner sep=0pt] at (0,-1.8) {     \includegraphics[width=0.48\textwidth]{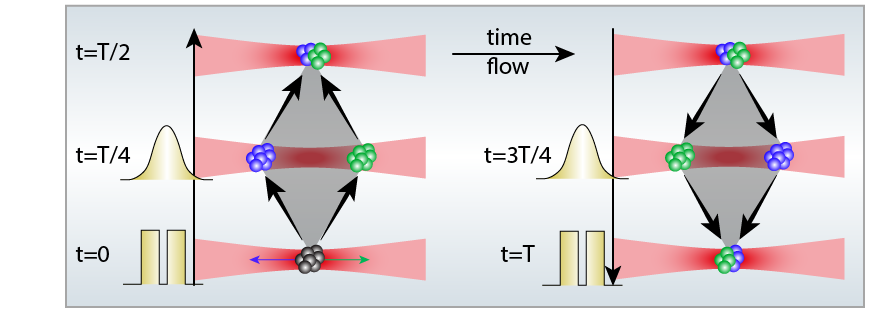}     };
\node[ align=left, rotate=-90] at (0.28,-4.3) {     \includegraphics[height=7.93cm]{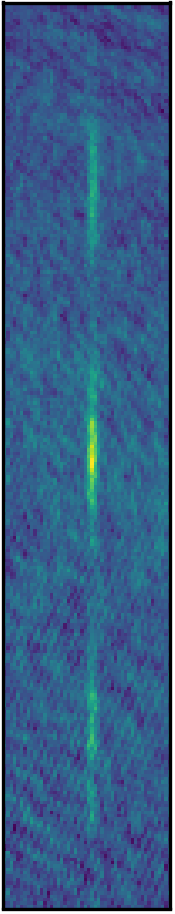}     };
\node[ align=left] at (-4.0, 2.2)     {a)};
\node[ align=left] at (-4.0, -0.6)     {b)};
\node[ align=left] at (-4.0, -3.7)     {c)};
\node[ align=left] at (2.7, -5.25)     {$p=2 \hbar k$};
\node[ align=left] at (0.4, -5.25)     {$p=0$};
\node[ align=left] at (-1.9, -5.25)     {$p= -2 \hbar k$};
\end{tikzpicture}
\caption{\label{fig:ExpScheme} a) Experimental setup. b) Interferometer sequence. Blue and green colors of atoms correspond to the $+ 2 \hbar k$ and $-2 \hbar k$ momentum components respectively. The time flow follows the black arrows. c) Experimental data acquired $\Delta t = \SI{12}{\milli \second}$ after the recombination pulse, with atoms in two channels:  $\ket{p=0}$ and $\ket{p=\pm 2 \hbar k}$.}
\end{figure}
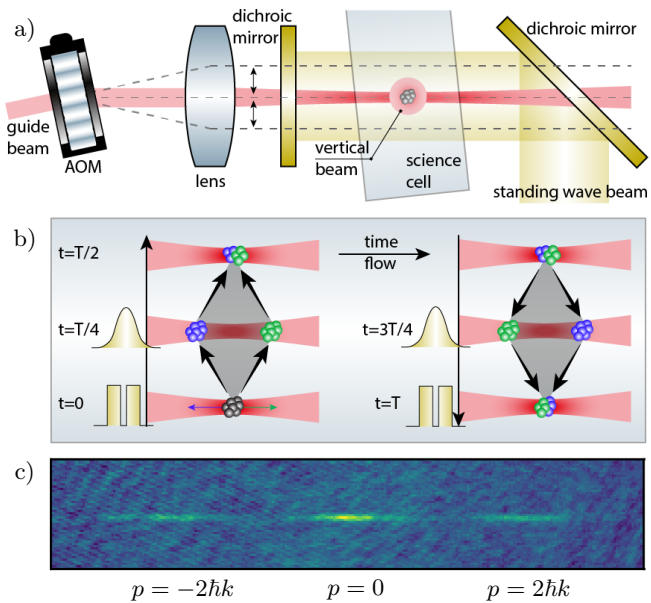

\section{Experimental Setup}
The BEC hardware is a modified version of the setup described in \cite{RyuNJoP2015}. A \textsuperscript{87}Rb BEC in the first-order magnetically insensitive state $ \ket{F=1, m_F = 0}$ and containing between 500 and 1500 atoms is prepared in a tight crossed-dipole trap formed at the intersection of a horizontal \SI{1064}{\nano \metre} laser beam with waist \SI{13}{\micro \metre} and a vertical \SI{1064}{\nano \metre} beam with waist \SI{60}{\micro \metre} and transverse trapping frequency $2\pi \times \SI{140}{\hertz}$. The \SI{60}{\second} cycle time of the experiment is dominated by the evaporative cooling stage in this trap. The interferometer waveguide is formed by a horizontal collimated \SI{1064}{\nano \metre} laser beam (Fig.~\ref{fig:ExpScheme}a) propagating collinear with the tight horizontal beam. The BEC is transferred adiabatically from the tight beam to the waveguide beam, still confined axially by the vertical beam. The waveguide power and waist are respectively \SI{550}{\milli \watt} and \SI{60}{\micro \metre}, giving axial and radial trapping frequencies of $2 \pi \times \SI{0.3}{\hertz}$ and $2\pi \times \SI{200}{\hertz}$. The waveguide beam can be translated horizontally over a few mm by the acousto-optic deflector (AOD) and lens shown in Fig.~\ref{fig:ExpScheme}a. It passes through the cell at a \SI{12}{\degree} angle of incidence to avoid undesirable interference resulting from multiple reflections at the cell windows.

Next, the confining vertical beam is switched off to allow the condensate to expand along the waveguide for \SI{20}{\milli \second}. The  wavepacket expansion which is undesirable during the interferometer cycle is then frozen with a \SI{1.5}{\milli \second} pulse of the vertical beam which realizes a delta-kick cooling potential \cite{ArnoldPRA2002}. The resulting wavepacket has an axial temperature of \SI{3}{\nano \kelvin} and an axial size of $\approx \SI{70}{\micro \metre}$ FWHM (see appendix A). The interferometer cycle starts immediately after the delta-kick pulse. The interferometer beamsplitters and mirrors are realized with pulses of a \SI{780}{\nano \metre} standing wave formed by retro-reflection of a beam with \SI{5.42}{\milli\metre} waist overlapped with the waveguide using two dichroic mirrors, as shown in Fig.~\ref{fig:ExpScheme}a.

As shown in Fig.~\ref{fig:ExpScheme}b, the BEC initially in momentum state $\ket{p=0}$ is split into a superposition of two wave packets with momentum $p= \pm 2 \hbar k$, where $k=2 \pi / \lambda$ using a double square pulse of the Bragg laser \cite{WuPRL2005, WangPRL2005,WuPRL2007,BurkePRA2009}. The two wave packets propagate in opposite directions with speed $v= \SI{11.7}{\milli\metre \per \second}$ while the waveguide is transversely moved out and back to the starting position \cite{WuPRL2007}. To create an interferometer with total interrogation time $T$, Gaussian reflection pulses \cite{muller2008atom} are applied at times $T/4$ and $3T/4$, followed by a second double square pulse which serves as a recombination pulse. During this process the atoms follow the diamond-shape trajectory shown in Fig.~\ref{fig:ExpScheme}b, enclosing a Sagnac area proportional to both $v T$ and the transverse displacement of the waveguide. The waveguide is translated with a time-dependent acceleration that minimizes the transverse excitation of the atoms at the final position \cite{TorronteguiPRA2011}. In our system this allows for transport of atoms up to \SI{2}{\milli \metre} in \SI{40}{\milli \second}.

The two output ports of the interferometer are the momentum states $p=0$ and $p = \pm 2 \hbar k$. The BEC evolves in the guide for \SI{10}{\milli \second} after the recombination pulse, allowing the momentum components to separate into three wavepackets easily detected via absorption imaging (in Fig.~\ref{fig:ExpScheme}c). For an interferometer with zero Sagnac phase, the population of atoms in each port $\left\{ N_{p=0},N_{\pm 2 \hbar k} \right\}$ depends on the interferometer phase $\phi$ as
\begin{equation}\label{eq:PopulationRatioVsPhase}
    P(\phi) = \frac{N_{p=0}}{N_{total}} = \frac{1}{2}  \left[  a \cos \left(  \phi + \phi_0 \right) + C  \right]
\end{equation}
where $N_{total}$ is the total atom number in the three ports, and $a$, $C$, and $\phi0$ are respectively the amplitude, offset, and phase shift of the fringe. The interferometer phase $\phi$ is
\begin{equation}
    \phi = 2 \left(     \phi_1 - 2 \phi_2 + 2\phi_3 - \phi_4    \right)
\end{equation}
where $\phi_i$ is the phase of the $i$th Bragg pulse [ Fig.~\ref{fig:ExpScheme}b]. The interferometer phase $\phi$ can be experimentally adjusted by changing the frequency $f$ of the Bragg laser by $\Delta f$ right before the recombination pulse, which changes the interferometer phase by
\begin{equation}
    \Delta \phi =    8 \pi L \Delta f / c 
\end{equation}
where $L=\SI{12.3}{\centi \metre}$ is the distance from atoms to the retro-reflector. This approach can scan $\phi$ over at least $4\pi$, as shown in Fig.~\ref{fig:fringesComparison}. 

\section{Results}
The interferometer has been operated in two modes: a static waveguide interferometer enclosing zero Sagnac area and the rotation-sensitive configuration obtained with the moving waveguide. The static mode aids optimization and characterization by decoupling any effects caused by waveguide movement from fundamental limitations of the system.
\begin{figure}
\includegraphics[width=0.48\textwidth]{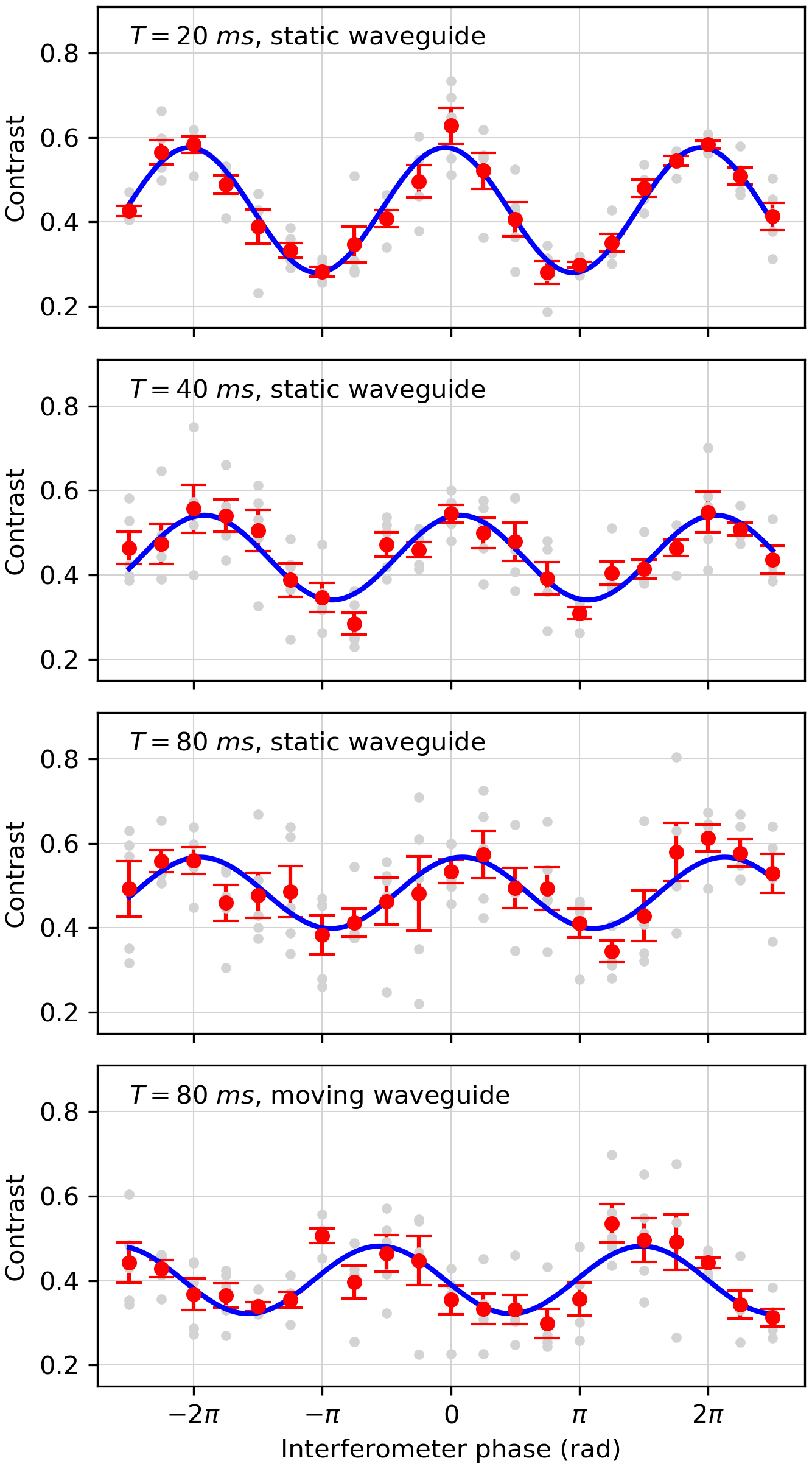}
\caption{\label{fig:fringesComparison} Interferometer fringes for a static waveguide atom interferometer with interrogation times $T =$ \numlist{20;40;80} \SI{}{\milli \second} and for a moving waveguide interferometer with $T = \SI{80}{\milli \second}$. Gray points represent the raw data for a single experimental run. Each red data point is an average of five repetitions for the given interferometer phase. Error bars are the standard error in those mean values.}
\end{figure}
Figure~\ref{fig:fringesComparison} shows interference fringes recorded for interrogation times $T=\SI{20}{\milli \second}$, \SI{40}{\milli \second} and \SI{80}{\milli \second} in the static mode, along with one moving guide fringe with $T=\SI{80}{\milli \second}$. The full $\pm \frac{5}{2} \pi$ interferometer scan was obtained by tuning the frequency of the final pulse over $\pm \SI{0.75}{\giga \hertz}$. Five measurements of $P(\phi)$ data at each phase are averaged and fitted to the function in Eq.~\ref{eq:PopulationRatioVsPhase}. In this data set, the moving guide encloses \SI{0.8}{\milli \metre \squared}, which is the largest Sagnac area enclosed by a fully guided atom interferometer reported to date. 

Figure~\ref{fig:fringesComparison} shows that there is a phase shift $\Delta \phi \approx \SI[parse-numbers=false]{0.5 \pi}{ \radian}$ between the moving and static guide interferometers for \SI{80}{\milli \second} interrogation times. Since both experiments were conducted under the same conditions this phase shift is presumably caused by the waveguide movement. Possible mechanisms include misalignment of the Bragg beam with the guide beam or center of mass motion excited by intensity fluctuations if the waveguide is tilted. While both mechanisms will be investigated and minimized in future work, the phase shift should not impact rotation sensing performance because the rotation rate can be obtained through a differential measurement with multiple Sagnac areas \cite{DurfeePRL2006}.

\begin{figure}
\includegraphics[width=0.48\textwidth]{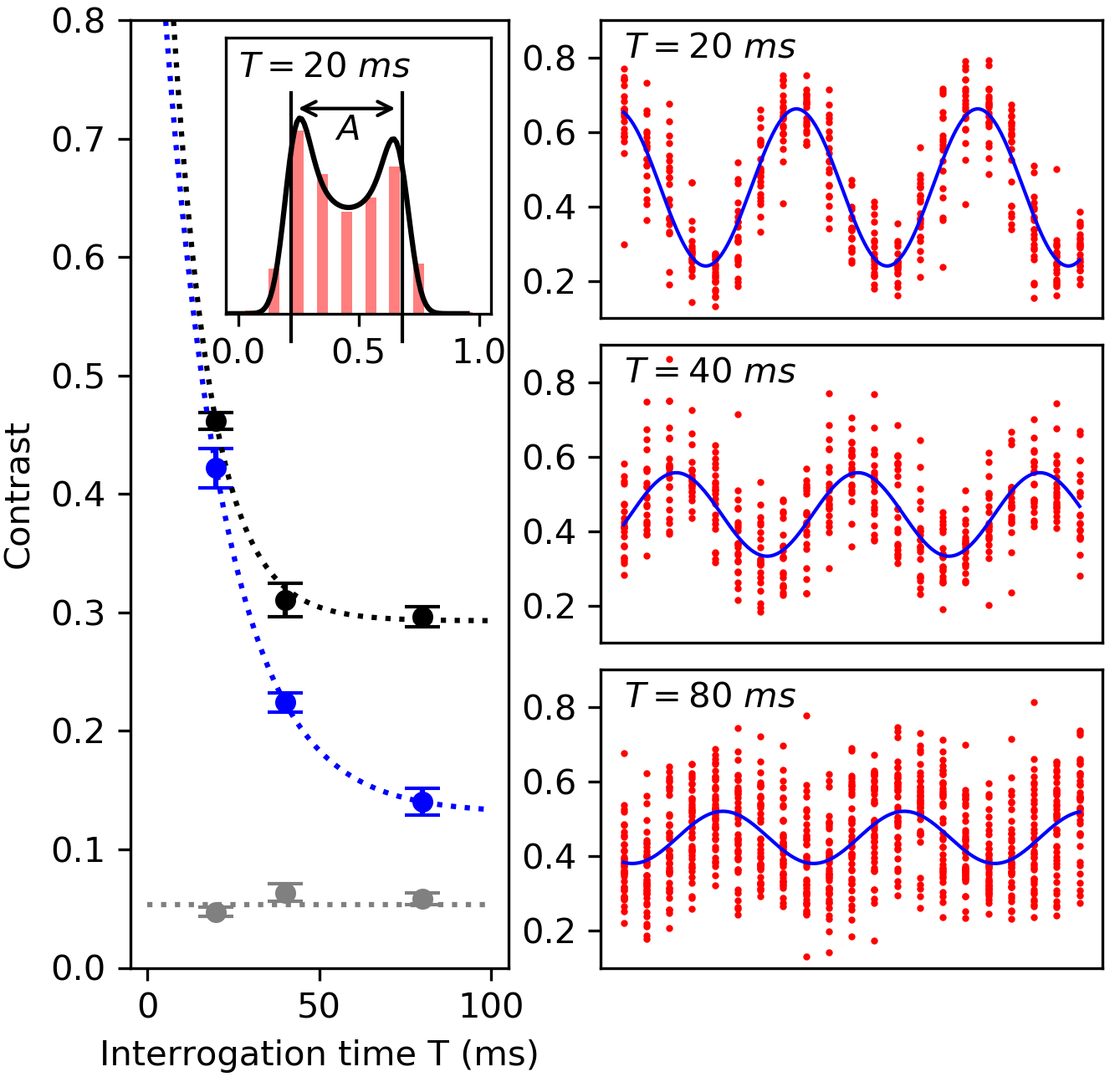}
\caption{\label{fig:ContrastAnalysis1} Interferometer fringe contrast as a function of interrogation time $T$. (Right) Each interferometer scan was repeated 22 times for $T= \SI{20}{\milli \second}$ and \SI{40}{\milli \second}, and 43 times for $T= \SI{80}{\milli \second}$. The blue sine curves are unweighted fits of Eq.~\ref{eq:PopulationRatioVsPhase} to the data. (Left)  The blue data points show the amplitude of the sinusoidal fits as a function of interrogation time $T$. The underlying contrast (black points) and the offset noise (gray points) were obtained by fitting the noise model discussed in the text to each phase of the fringe. The dotted curves are to guide the eye. The inset plot shows a histogram of all data points in the  $T= \SI{20}{\milli \second}$ fringe. The distance A is the underlying contrast of the fringe.}
\end{figure}

The behavior of both the underlying and average fringe contrast with interrogation time evident in Fig.~\ref{fig:fringesComparison} can provide useful information about imperfections and noise processes in the interferometer. Factors impacting the underlying contrast have been extensively studied in \cite{BurkePRA2008}, which showed that a residual axial curvature of the approximately collimated guide imparts a spatially-varying phase shift to the wave packets that can lead to decreased fringe visibility. The analysis showed that this effect can be minimized by ensuring the interferometer is symmetrical, wavepacket size remains constant, and the timing of the Bragg pulses is optimized. The delta-kick cooling technique employed in our interferometer helps in this regard because it minimizes wavepacket size changes during the interferometer cycle. Residual axial curvature also decreases the time at which the wavepackets overlap during the recombination pulse by an amount $\delta T \approx 2 \omega^2_{a} (T/4)^3$ in a waveguide with axial frequency $\omega_a$ \cite{BurkePRA2008}. For our interferometer with $\omega_{a}=2 \pi \times \SI{0.3}{\hertz}$ and total interrogation time $T=\SI{80}{\milli \second}$ the timing shift is $\delta T \approx \SI{0.06}{\milli\second}$, which is negligible here. In future experiments with longer interrogation times the effect will become important, requiring that the Bragg pulse timing be adjusted accordingly. 

\subsection{Noise analysis}
Since a detailed study of the fringe contrast supported by an appropriate noise model might reveal factors limiting the system, static mode fringes for our standard set of $T=\SI{20}{\milli \second}$, \SI{40}{\milli\second} and \SI{80}{\milli \second} with many repetitions per interferometer phase were collected over several hours to produce a large data set for further analysis (Fig.~\ref{fig:ContrastAnalysis1}). The raw data presented in Fig.~\ref{fig:ContrastAnalysis1} right column (red points) shows that the range of $P(\phi)$ values is roughly the same for all three interrogation times, while the fluctuations of $P(\phi)$ at each phase increase with interrogation time. There are two intrinsic noise sources for all atom interferometers: shot noise and phase diffusion \cite{JoPRL2007}. Shot noise is negligible compared to the noise visible seen in the fringes presented here. An estimate of the reduction in average contrast due to phase diffusion \cite{LewensteinPRL1996, JavanainenPRL1997,LeggettPRL1998_comm, JavanainenPRL1998_res, BistritzerPNAS2007, BurkovPRL2007, HofferberthNature2007, IloOkekePRA2010, GrondNJoP2010, StimmingPRA2011, FallonJoPB_AMOP2015} says that the average contrast will still be 90\% of the maximum value at $T=$\SI{80}{\milli \second} (see Appendix C). So it appears that the current interferometer is limited by technical noise.

Factors that could cause fluctuations in $P(\phi)$ (Eq.~\ref{eq:PopulationRatioVsPhase}) include noise in the interferometer phase $\phi$, noise in the underlying contrast $a$, and noise in the offset $C$. A known source of technical noise in our experiment is mechanical vibrations of the Bragg retro-reflector mirror that cause changes in the phase of the Bragg pulses at the location of the atoms: If the optical table supporting the experiment is not floated the resulting phase noise causes all measurements to average to the same value within the error bars for all phases $\phi$, even for our shortest interferometer time $T=\SI{20}{\milli \second}$.

\begin{figure}
\includegraphics[width=0.48\textwidth]{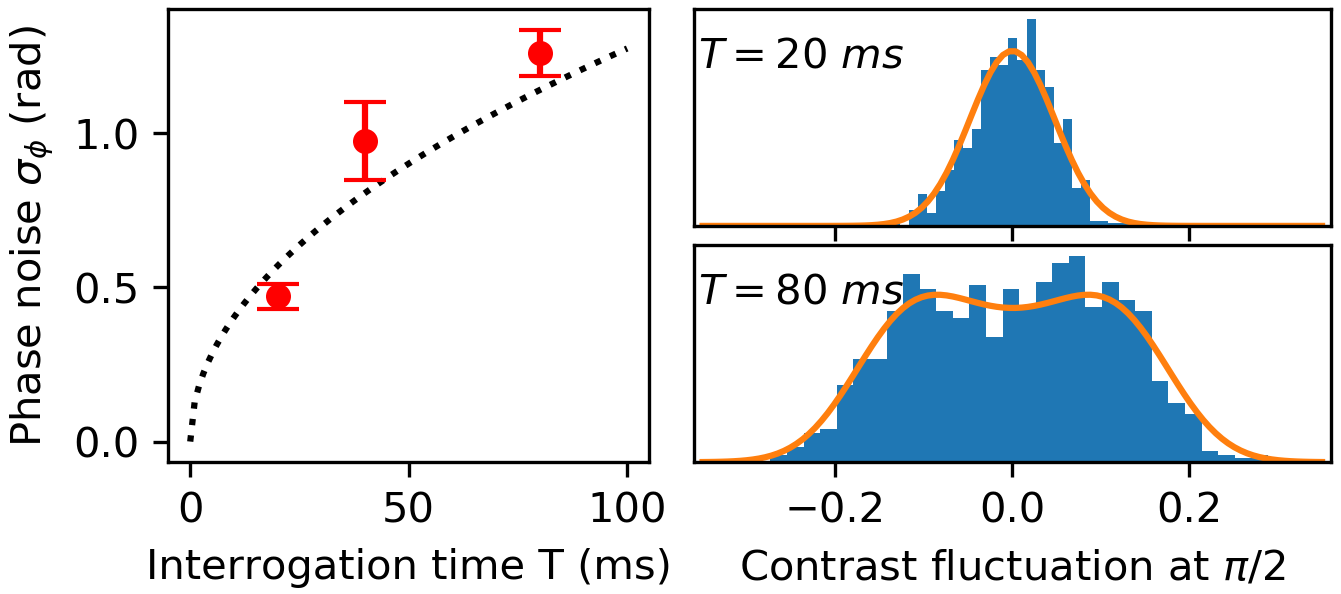}
\caption{\label{fig:ContrastAnalysis2} Noise dependence on the interrogation time $T$. (left)  Standard dev of the phase noise $\sigma_\phi$ of the contrast values distribution $P_i$  at the steep side of the fringe $ \phi =  {\pi}/{2}$ . The fit is $c\sqrt{T}$. (right) the histogram of the atoms fraction distribution minus the average value of the full set $P_i-\bar{P}$. The noise model with standard deviation $\sigma_\phi$ determines the shape of the orange curves. Both histograms consist of about 800 samples total.}
\end{figure}

The shape of a histogram of the $P(\phi)$ values recorded for multiple scans over the fringe (inset in Fig.~\ref{fig:ContrastAnalysis1} left column) provides information about sources of noise in $P(\phi)$. Pure phase noise does not change the shape of the histogram from the harmonic distribution of a noiseless interferometer. So the blurred edges of the histogram in Fig.~\ref{fig:ContrastAnalysis1} suggest that there is at least one other source of technical noise. The noise model has three free parameters: underlying contrast $A$, phase noise $\sigma_\phi$ and offset noise $\sigma_c$. With sufficiently good statistics it becomes possible to determine the dependence of those three contributions on interrogation time $T$ (see Appendix B).

The decreasing amplitude $a(T)$ of the fitted fringes (blue curves) versus the interrogation time is shown in the left part of Fig.~\ref{fig:ContrastAnalysis1} as blue points. The decreasing trend is a result of averaging over all possible sources of noise. The underlying amplitude $A$ in the fitted noise model (black points) also decreases with interrogation time. At this stage we can only speculate on what causes the decrease in the underlying contrast, but it is possible that the experiment was just not as well optimized for $T=\SI{40}{\milli \second}$ and $T=\SI{80}{\milli\second}$ as it was for $T=\SI{20}{\milli \second}$. 

The noise model analysis of the fringes shows that the offset noise is constant (gray points in Fig.~\ref{fig:ContrastAnalysis1} left), while the phase noise increases significantly with interrogation time. The phase noise $\sigma_\phi$ obtained from the noise model is plotted against the interrogation time in Fig.~\ref{fig:ContrastAnalysis2} (left) and appears to scale $\propto \sqrt{T}$. Figure~\ref{fig:ContrastAnalysis2} (right) shows the distribution of the measured atom population $P(\phi)$ at the steepest point of the fringe ($\phi = \pi /2$). The distribution for $T=\SI{20}{\milli \second}$ is Gaussian whereas the distribution for $T=\SI{80}{\milli \second}$ has two visible peaks away from the mean value. This is consistent with the noise analysis and indicates that the Gaussian phase fluctuation grows with time and exceeded $\pi /4$ at $T=\SI{80}{\milli \second}$. The conclusion obtained from this analysis is that,since the offset noise remains constant, the dominant source of noise for interferometer interrogation times longer that \SI{20}{\milli \second} is mechanical vibrations of the retro-reflection mirror. 

While the unknown source of offset noise makes removing it difficult without further in-depth studies, the phase noise can be removed using the technique presented in \cite{GeigerNatureCommunications2011}. Here, the performance of an atom interferometer operated in the especially harsh environment of an aircraft in flight was greatly improved by mounting a high-performance accelerometer on the retro-reflector and thereby correcting for vibration-induced changes in its position during the interferometer cycle. In addition, engineering better mounting solutions for the retro-reflector will also improve interferometer performance.

\subsection{Long-term stability}

\begin{figure}
\includegraphics[width=0.48\textwidth]{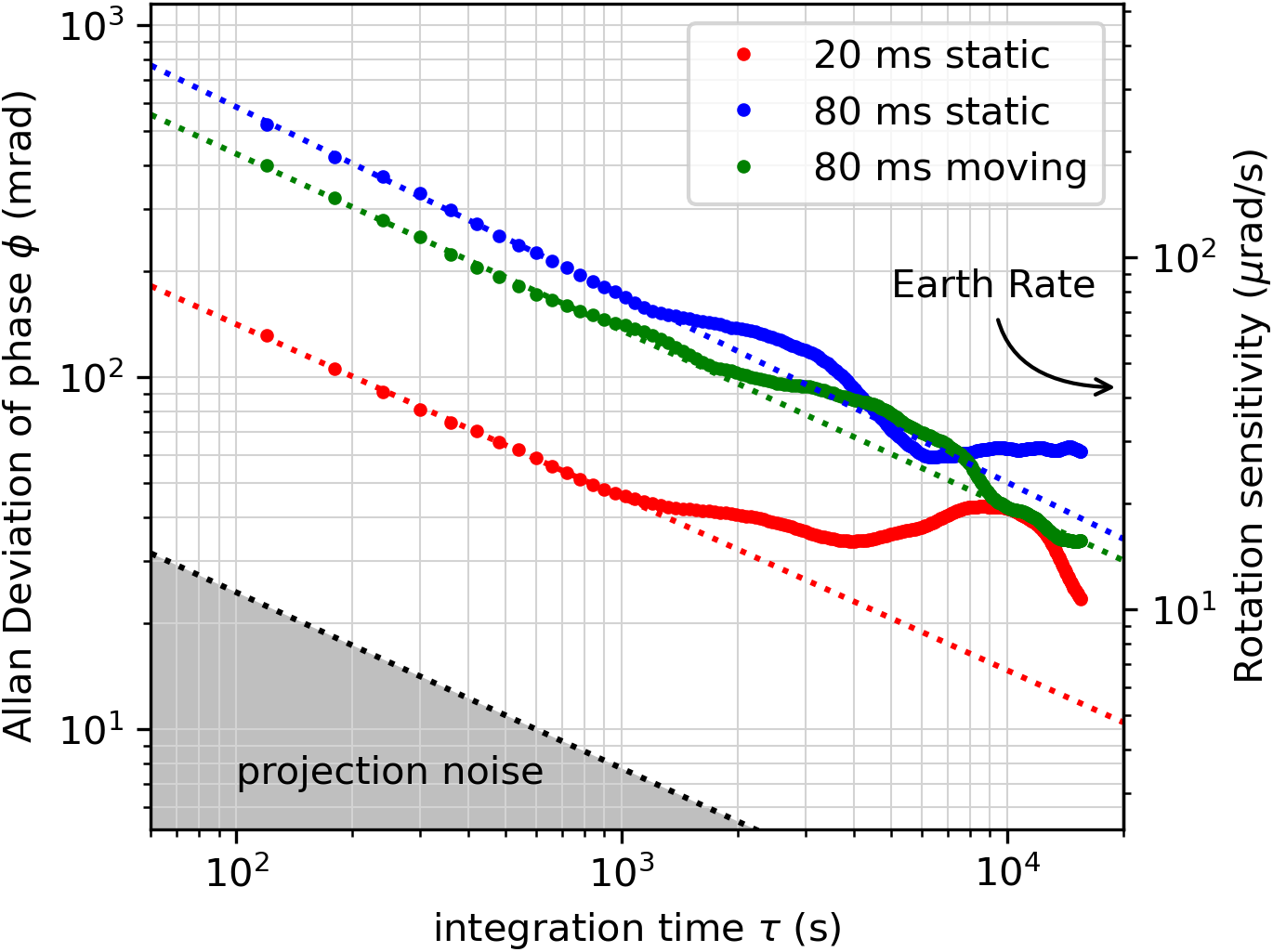}
\caption{\label{fig:AllanDev} Overlapping Allan deviation scaled by the corresponding underlying contrast for different interferometer cycle times. The projection noise is computed as $1/a \sqrt{N} $ for $a=1$ and $N=1000$. The right hand axis shows the corresponding rotation sensitivity of the $T=\SI{80}{\milli \second}$ moving guide interferometer which has \SI{0.8}{\milli \metre \squared } enclosed area. The arrow labeled Earth Rate corresponds to a projection of the Earth's rotation vector onto the horizontal loop area in New Mexico.}
\end{figure}

The capacity to average measurements for a long time is key for precision measurement. Accordingly, the long-term stability of our system has been assessed by measuring the Allan deviation \cite{NgJofGCN1997} of the interferometer phase at the steepest point of the fringe over 12 hours for three cases: static \SI{20}{\milli \second} interferometer, static \SI{80}{\milli \second} interferometer and moving \SI{80}{\milli \second} interferometer. Results are presented in Fig.~\ref{fig:AllanDev} for integration times $\tau$ up to 5.5 hours. The \SI{60}{\second} cycle time of the experiment determines the minimum integration time $\tau = \SI{120}{\second}$. The plot shows that the Allan deviation scales $\propto 1 / \sqrt{\tau}$ for all three cases. The extrapolated values at $\tau = \SI{1}{\second}$ are \SI{1300}{\milli \radian} and \SI{2600}{\milli \radian} for $T=\SI{20}{\milli \second}$ and $T=\SI{80}{\milli \second}$ respectively. The Allan deviations for stationary and moving \SI{80}{\milli \second} interferometer are in good agreement, indicating that moving the waveguide does not influence the stability of the device even though a phase shift is induced by translating the waveguide. The Allan deviation measurement shows that the sensitivity after \SI{2000}{\second} ($\approx\SI{30}{\min}$) of averaging is comparable to the Earth's rotation rate (\SI{42}{\micro \radian \per \second } in New Mexico) for the $T=\SI{80}{\ms}$ moving waveguide Sagnac interferometer.

\section{Discussion and Conclusions}
In summary, we have demonstrated the first BEC matter wave analog of a fiber optic gyro. It encloses largest Sagnac area (\SI{0.8}{\milli \metre \squared}) achieved for a waveguide device. The total interrogation time was extended up to \SI{80}{\milli\second} by implementing a form of delta-kick cooling to reduce the BEC temperature, generating dilute wave-packets that remain collimated throughout the interferometer cycle and suppressing phase diffusion. The average fringe contrast decreases with interrogation time, primarily due to phase fluctuations produced by mechanical vibrations of the retro-reflector forming the standing wave for the Bragg pulse beamsplitters and mirrors. Statistical analysis shows that the dominant noise source is Gaussian (white power spectrum), with the Allan deviation falling with averaging time $\tau$ as $\tau^{-1/2}$ for $\tau$ up to 10,000 seconds.

The architecture of our guided Sagnac atom interferometer offers some advantages over other approaches.
In addition to increasing the Sagnac area, the use of two reflections instead of one creates a symmetry which suppresses contrast loss due to residual axial curvature \cite{BurkePRA2008} and also makes the device insensitive to constant acceleration along the guide. The interferometer is relatively insensitive to changes in the waveguide laser intensity; while the guide beam power was stabilized to $\approx1 \%$ before the AOM, the guide beam power in the cell varied systematically by about 20\% with guide movement due to the AOM's frequency response. It is known that the mode of the beam might be distorted by an AOM. While not important here, these issues will be addressed in future work with increased Sagnac area by stabilising the guide power after the science cell, and by replacing the AOM with a piezo-controlled mirror.

In most guided AI designs, measuring rotation along a horizontal axis entails turning the device on its side, which will require some form of gravity compensation to nullify the effects of gravity on the now vertical wavepacket trajectories. An advantage of our approach is that replacing our current 1D AOD with a 2D AOD and a folding mirror will enable waveguide translation along all three axes and hence three-axis measurement with a single fixed-orientation device without gravity compensation. Although it is possible to engineer a device performing measurements along all axes simultaneously, by for instance combining three independent sensors, the most straightforward implementation with the current setup is to perform measurement along each axis sequentially. This will decrease the bandwidth of our sensor by a factor of 3, but it will not affect the performance of the device otherwise. Moreover, the tight transverse confinement in our FOG analog ($\approx \SI{200}{\hertz}$) will provide good protection against platform motion. In comparison, multi-axis measurement may be more challenging in architectures based on wavepackets orbiting in weak ($\approx \SI{2}{\hertz}$) bowl-like potentials, where both platform motion and tilting the potential would result in asymmetric trajectories that degrade performance \cite{MoanPRL2020}. Moreover, the bandwidth of our interferometer, where a single measurement is a direct measurement of Sagnac phase, should be higher than that of differential devices \cite{MoanPRL2020} where phase determination requires reconstruction of a relative phase ellipsoid comprised of several interferometer cycles. The statistical sensitivity demonstrated in this work will make possible a precise measurement of the rotation rate of the Earth within 30 minutes using a differential scheme with multiple Sagnac areas.

It should be possible to improve the performance of our device by several orders of magnitude \cite{NarducciAdvInPhys2022}. Future work will focus on increasing the Sagnac area to increase the sensitivity to rotation. The current experimental limitations on area are the trap depth of the moving guide and the AOD deflection range, which together limit the maximum distance over which atoms can be translated within the coherence time. It will be possible to increase both the waveguide trap depth and the distance over which the guide is moved by changing the hardware. Also, since the interferometer phase noise is dominated by vibrations of the retro-reflecting mirror forming the standing wave, adding a suitable accelerometer to the retro-reflector \cite{GeigerNatureCommunications2011} and engineering that part of the system to improve its stability will increase the coherence time. That will in turn allow us to demonstrate multiple round-trips in the waveguide loop \cite{kim2022one}, which is a major advantage of the waveguide geometry. We will also increase the wavepacket momentum using optimized Bragg pulse shapes \cite{cassidy2021improved}, which will further increase the area enclosed in a given interrogation time. For example, with wavepacket momentum $10\hbar k$, implementation of 200 round-trips as demonstrated in \cite{kim2022one}, and higher laser power of 100W allowing for roughly 10 times further guide translation due to increased radial trap frequency, the expected enclosed area will be \SI{80}{\centi\metre\squared}. The combination of this increased area with the noise level of the current device would lead to an angular random walk (ARW) of \SI{0.0004}{\degree \per \sqrt{\hour}}, which is similar to that of state of the art FOGs \cite{KorkishkoConf2013,KorkishkoIEEEsymposium2018}.

\section{Acknowledgements}
We thank Dana Anderson for helpful discussions of his previous work on waveguide Sagnac atom interferometers. This  work  was  funded  by DARPA under the A-PhI program. The initial development of the experimental apparatus was supported by the Office of Naval Research and by the Laboratory Directed Research and Development program of Los Alamos National Laboratory under project number 20180045DR.

\section{Appendix A. $\delta$-kick cooling}

$\delta$-kick cooling (DKC) \cite{ArnoldPRA2002} applies a pulsed harmonic confining potential to a condensate to manipulate its momentum distribution. In this work DKC is applied after the BEC is prepared in a crossed-dipole trap formed of two perpendicular  \SI{1064}{\nano \metre} beams having waist $\omega_0=\SI{60}{\micro \metre}$: a horizontal guide beam and a vertical beam. Here the vertical beam provides tight confinement of the BEC along the guide beam axis.

After the vertical beam is switched off, the condensate expands freely along the waveguide for \SI{20}{\milli \second} before the vertical beam is pulsed on for \SI{1.6}{\milli \second} to provide the DKC confining potential. The size of the cloud along the guide $\sigma_{a}$ is obtained by fitting absorption images with 2D Gaussian function:
\begin{equation}
  g(x,y) = \exp \left[ - \frac{\left( x-x_0 \right)^2 }{2 \sigma^2_{a}} - \frac{\left( y-y_0 \right)^2 }{2 \sigma^2_r} \right]
\end{equation}
where $\sigma_r$ is size in the radial direction of the guide, and $(x_0, y_0)$ is the position of the center of the cloud.

\begin{figure}
\includegraphics[width=0.48\textwidth]{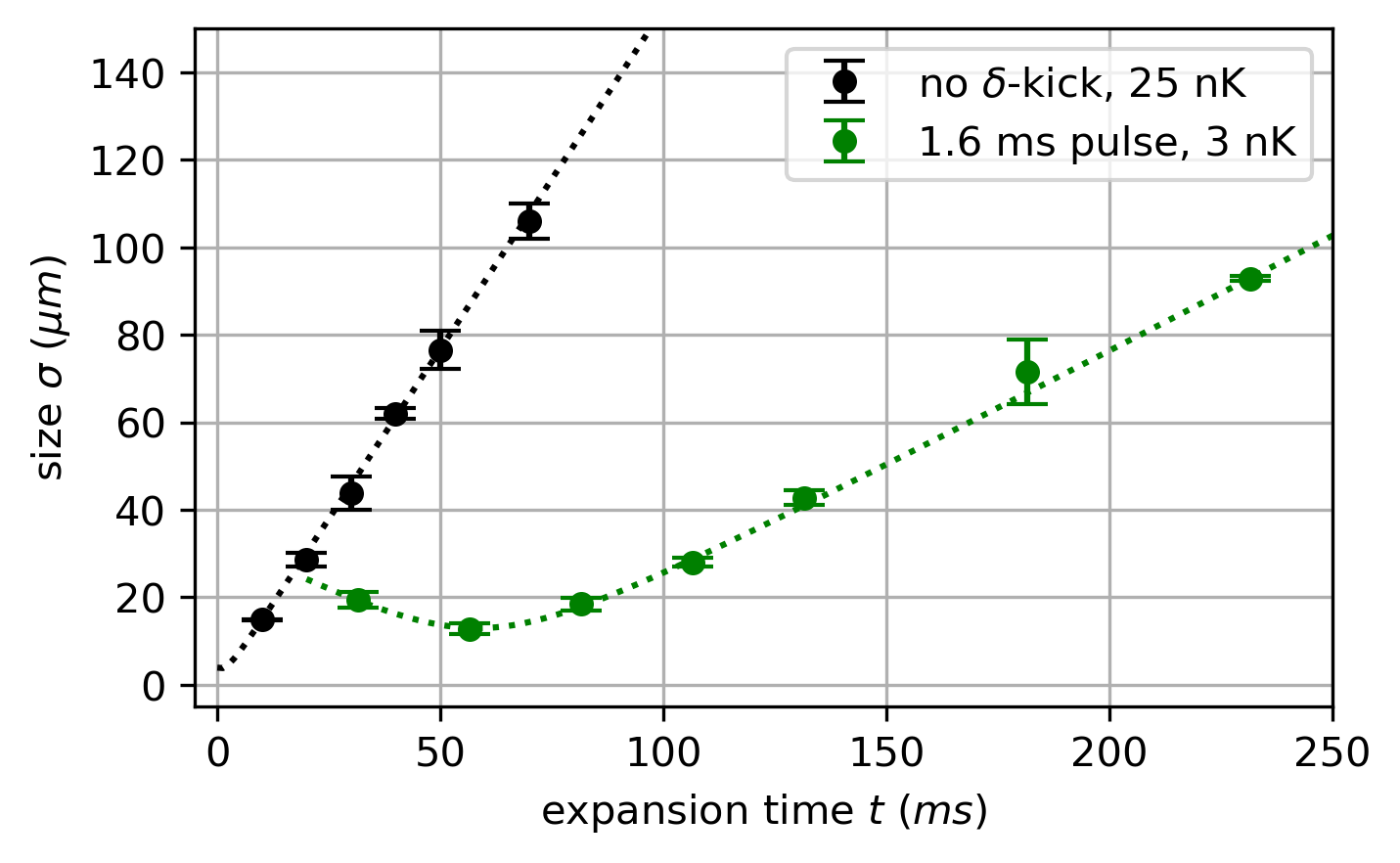}
\caption{\label{fig:AppendixDeltaKickCooling} BEC size $\sigma_{a}$ along the guide for variable expansion time $t$. Each data point is a mean of three measurements with an error bar given by the standard error in the mean. The dotted lines are a fit of equation \ref{eq:TempExpanstion} to the data, finding temperatures of \SI{25\pm 3 }{\nano \kelvin} without $\delta$-kick cooling, and \SI{2.95\pm 0.08 }{\nano \kelvin} with a \SI{1.6}{\milli\second} $\delta$-kick pulse. }
\end{figure}

The cloud expansion with and without DKC as a function of expansion time $t$ is illustrated at figure \ref{fig:AppendixDeltaKickCooling}. The data is fitted with a simple thermal expansion model 
\begin{equation}\label{eq:TempExpanstion}
    \sigma(t) = \sqrt{ \sigma^2_{t=0} + \frac{k_B T_{BEC}}{m_{Rb}} t (t-2 t_f) }
\end{equation}
to obtain the temperature $T_{BEC}$ of the condensate, where $m_{Rb}$ is the rubidium mass, $k_B$ is Boltzmann's constant, $\sigma_{t=0}$ is the size of the cloud at $t=0$, and $t_f$ the time at which focusing of the BEC occurs. When no delta kick cooling is applied, the cloud expands rapidly with temperature $\approx \SI{25}{\nano \kelvin}$. With DKC the cloud is weakly focused at $t_f = \SI{38}{\milli \second}$ after the DKC pulse. The much slower subsequent expansion with DKC  corresponds to a temperature of \SI{3}{\nano \kelvin}.

The final temperature delivered by our DKC implementation is currently limited by technical factors. We find that a pulse duration of \SI{1.6}{\milli \second} results in the lowest temperature and the most symmetric size evolution over a \SI{80}{\milli\second} interferometer interrogation time. 
Increasing the duration of the DKC pulse beyond this does not decrease the temperature further, because the expansion dynamics are predominantly governed by the mean-field energy of the condensate. The expansion time before the DKC pulse is limited by the size of the vertical beam. The size of the cloud after initial expansion of $\SI{20}{\milli \second}$ reaches $\text{FWHM}_{BEC} \approx \SI{70}{\micro \metre}$, which is comparable to the size of the beam. Increasing both the power and the size of the vertical beam should lead to even lower temperatures.

\section{Appendix B. Noise model}

Here we present a noise model describing the histogram of $P(\phi)$, the $p=0$ fraction in the interferometer output ports.  
It extends the model containing only normally-distributed offset noise used in \cite{GeigerNatureCommunications2011} to extract information about readout noise. Our model contains two normally-distributed variables: interferometer phase $\phi$ with standard deviation $\sigma_{\phi}$ and an additive offset noise with standard deviation $\sigma_c$. Even though this case study is applied to interferometer data, the model also holds for any type of data described by sinusoidal behaviour. 

Let us first recall that the fraction of atoms in the $p=0$ interferometer output port is
\begin{equation} \label{eq:Fraction}
    P(\phi) = \frac{N_{p=0}}{N_{total}} = \frac{1}{2}  \left[  a \sin \left(  \phi - \frac{\pi}{2} \right ) + C  \right],
\end{equation}
where $a$ is the amplitude of the fringe.

\begin{figure*}
\includegraphics[width=0.98\textwidth]{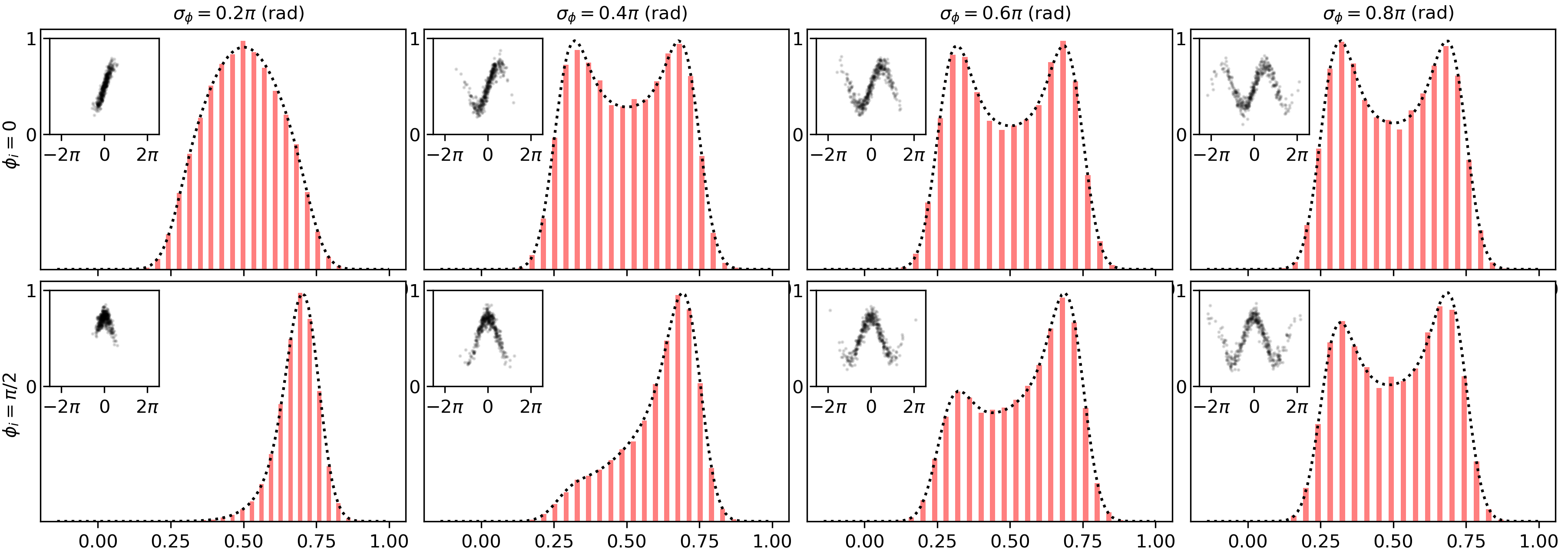}
\caption{\label{fig:AppendixNoiseModel2} The influence of phase and offset noise on the output of the interferometer. Synthetic data was obtained by adding normally-distributed phase and offset noise to the interferometer fringe function Eq.~(\ref{eq:Fraction}) for fixed values of $\phi_i =0$ and $\phi_i=\frac{\pi}{2}$. The offset noise $\sigma_c = 5\%$ remains the same for all the graphs, while the phase noise increases from left to right. Inserts show how the phase fluctuation explores the fringe. The black dotted line is the noise model calculated for given values of $\phi_i$, $\sigma_\phi$ and $\sigma_c$ using Eq.~(\ref{eq:HistogramModel}).}
\end{figure*}

In the first step we will derive the model for phase noise only, assuming $C=0$ and neglecting $\pi/2$ term for simplicity. This means that the distribution is centered around 0. If a probability density function (PDF) of the phase $\phi$ is known, then one can use it to derive a PDF of the atom fraction $P$. Following this argument, the total probability of the measured data reflects the probability of the noisy variable $PDF(P)dP = PDF(\phi)d\phi$. Since the contributions from the normal distribution of phase $\phi$ can exceed $\pi/2$, the formula for $PDF(\phi)$ takes the form 
\begin{equation}\label{eq:PDF}
PDF(\phi) = \frac{1}{\sqrt{2 \pi } \sigma_\phi } \sum_{n=-\infty }^{\infty} \exp{  \frac{(\phi - (-1)^n\phi_i - n\pi)^2 }{-2 \sigma_\phi^2}}
\end{equation}
where $\phi$ is limited to $<-\pi/2, \pi/2>$ and $\phi_i$ is the Bragg phase. Changing variables from $\phi$ to $P$ (using $\frac{d\phi}{dP}$ and $\phi(P) = \arcsin{(2P/a)}$) and combining it with the expression for $PDF(\phi)$, one finds
\begin{equation}
PDF(P) =2  PDF(\phi (P)) /  a \sqrt{ 1 - (2 P/a)^2  }
\end{equation}

Since the histogram of the data represents the probability that certain outcome is within a finite range of values, the equation above needs to be integrated to represent the quantity measured in the experiment accurately, giving
\begin{equation}\label{eq:PurePhaseNoiseHistModel}
    CDF(P) = \frac{1}{2} \sum_{n=-\infty }^{\infty} \text{erf} \left[ \frac{\arcsin{\frac{2 P}{a}} -(-1)^n \phi_i - n\pi }{\sqrt{2} \sigma_\phi}   \right]
\end{equation}
Now, Eq. (\ref{eq:PurePhaseNoiseHistModel}) evaluated for bins of the histogram $F(P)\rvert_{P_i}^{P_{i+1}} $ is an accurate representation of the histogram data. In the following step the additive independent normal noise $\sigma_c$ can be introduced by calculating the convolution of the $F(P)$ and Gaussian distribution:
\begin{equation}\label{eq:HistogramModel}
    (F(P) * g(P)) = (F(P) * \frac{1}{\sqrt{2 \pi} \sigma_c} \exp{ - \frac{  P ^2}{ 2 \sigma_c^2} } )
\end{equation}
The convolution is easily calculated numerically, allowing the model to be implemented to analyze the data in this paper. Figure~\ref{fig:AppendixNoiseModel2} shows how the histogram shape for measurements depends on the noise. The important feature of the model is that phase and offset noise are orthogonal. Moreover, even for a substantial phase noise close to $\sigma_\phi < \pi$, where values of the phase explore more than one fringe, the model can be used to determine the exact value of phase noise spread $\sigma_\phi$.

\begin{figure}
\includegraphics[width=0.48\textwidth]{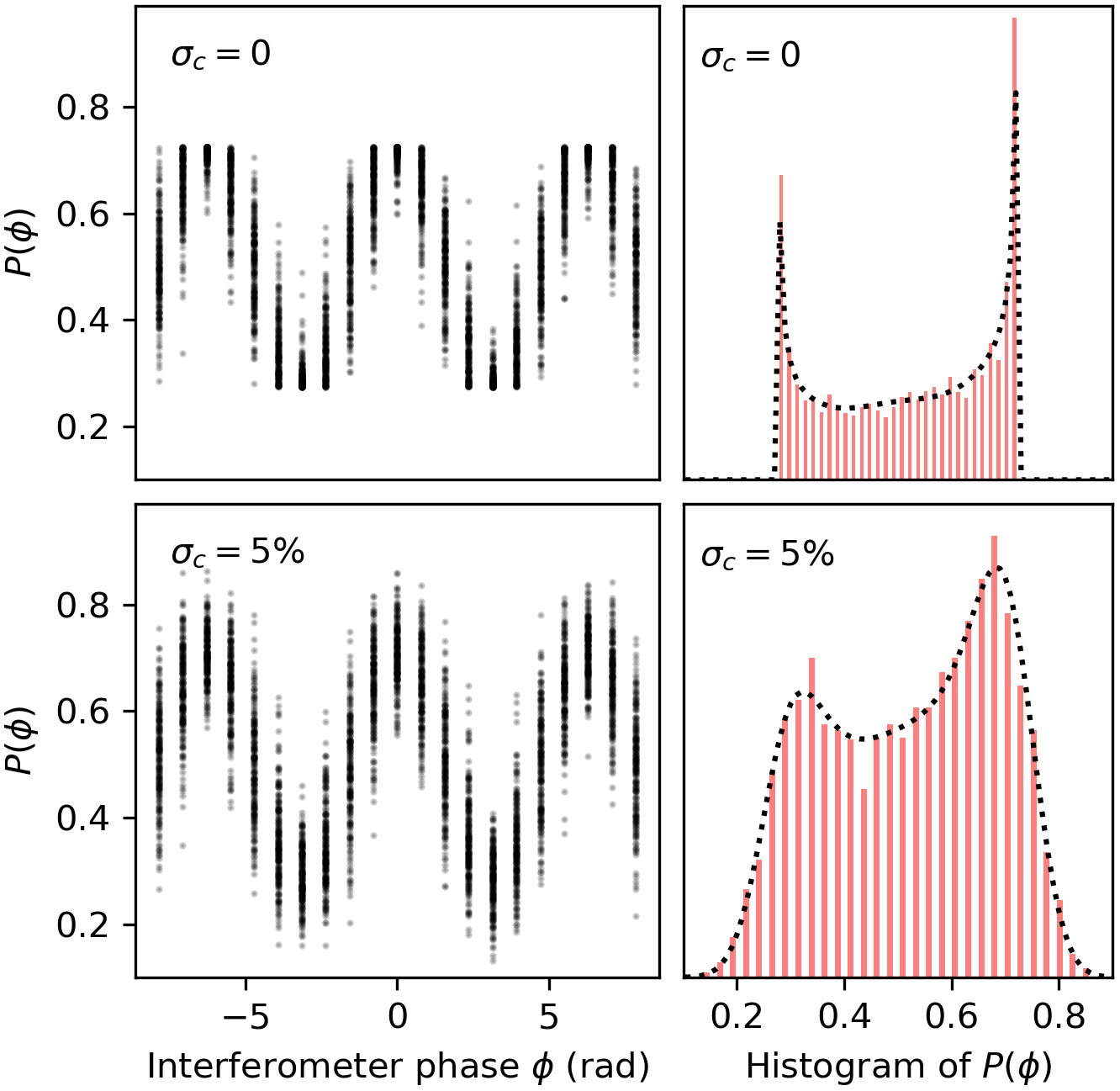}
\caption{\label{fig:AppendixNoiseModel1} Simulated interferometer data and the corresponding histograms of $P(\phi)$, the relative atom population in the $p=0$ port. Blue curves show the interferometer fringe with no noise. The individual data points were obtained using Eq.~(\ref{eq:Fraction}), with fixed amplitude $a=0.45$ and normally-distributed phase noise around $\phi_0=0$ [rad] with standard deviation $\sigma_\phi = 0.4$ [rad]. The upper row shows the fringe and the histogram for no offset noise $\sigma_c = 0$, while the offset noise $\sigma_c=0.05$ [rad] is introduced in the bottom row. The dotted lines represent the noise model for each histogram obtained from Eq.~(\ref{eq:FullFringeNoiseModel}).}
\end{figure}

If the experimental data is obtained by repeating the measurement for a  series of discrete interferometer phases $\left[ \phi_1, \phi_1, \dots, \phi_n \right]$, then the histogram of the full data set is simply a sum of Eq.~(\ref{eq:HistogramModel}) over the entire set,
\begin{equation}\label{eq:FullFringeNoiseModel}
    \sum_{\phi_i}  (F(P, \phi_i) * g(P)). 
\end{equation}
Figure~\ref{fig:AppendixNoiseModel1} (right column) shows a comparison between the model and a synthetic data set obtained by adding normally distributed noise to a cosine curve described by Eq.~(\ref{eq:Fraction}). As shown by the figure, the model (dotted black line) is in a good agreement with the synthetic data (histogram bars). The comparison clearly shows that the histogram has sharp edges for an interferometer where the phase fluctuation is a dominant source, while the edges are blurred when the contribution to noise from other sources is significant.

Even though the noise model presented here does not include noise in the amplitude, it can still help distinguish between phase noise, which is a typical source of noise in interferometers, and other sources of technical noise.

\section{Appendix C. Phase Diffusion}
Phase diffusion is known to be one of the intrinsic factors limiting the performance of atom interferometers. It has been observed and studied in various architectures \cite{LewensteinPRL1996, JavanainenPRL1997,LeggettPRL1998_comm, JavanainenPRL1998_res, BistritzerPNAS2007, BurkovPRL2007, HofferberthNature2007, IloOkekePRA2010, GrondNJoP2010, StimmingPRA2011, FallonJoPB_AMOP2015}. 
The diffusion rate for an approach where atoms are split and recombined nonadiabatically using Bragg pulses, similarly to the setup used in our experiment, was theoretically studied in \cite{IloOkekePRA2010}. Here, the effect of phase diffusion is considered to be created only by the fluctuations of the atom numbers in each cloud, which are governed by a binomial distribution.

Following the discussion in \cite{IloOkekePRA2010, FallonJoPB_AMOP2015}, the effect of the interatomic interactions in a many-body Hamiltonian is captured by the interaction coefficient $g$
\begin{equation}
g = \frac{U_0 }{2}\int  |\psi_\pm|^4 \mathbf{dr}
\end{equation}
where $U_0 = 4\pi \hbar^2 a_{sc}/M$, with $M$ being the atomic mass, $a_{sc}$ the s-wave scattering length, and $\psi_\pm$ are wavefunctions of two separate clouds after splitting. Here the wavefunction $\psi_\pm$ of each cloud is normalized to 1.
The accumulated nonlinear phase $\xi$ per atom is then obtained by integration of the interaction coefficient $g$ over interrogation time $T$:
\begin{equation}
\xi = \frac{1}{\hbar } \int_0^T dt g
\end{equation}
Knowing the accumulated phase $\xi$ and the total number of atoms $N$, the averaged fringe visibility $V$ is described by 
\begin{equation}
V = \exp{-2 \xi^2 N}
\end{equation}

\begin{figure}
\includegraphics[width=0.48\textwidth]{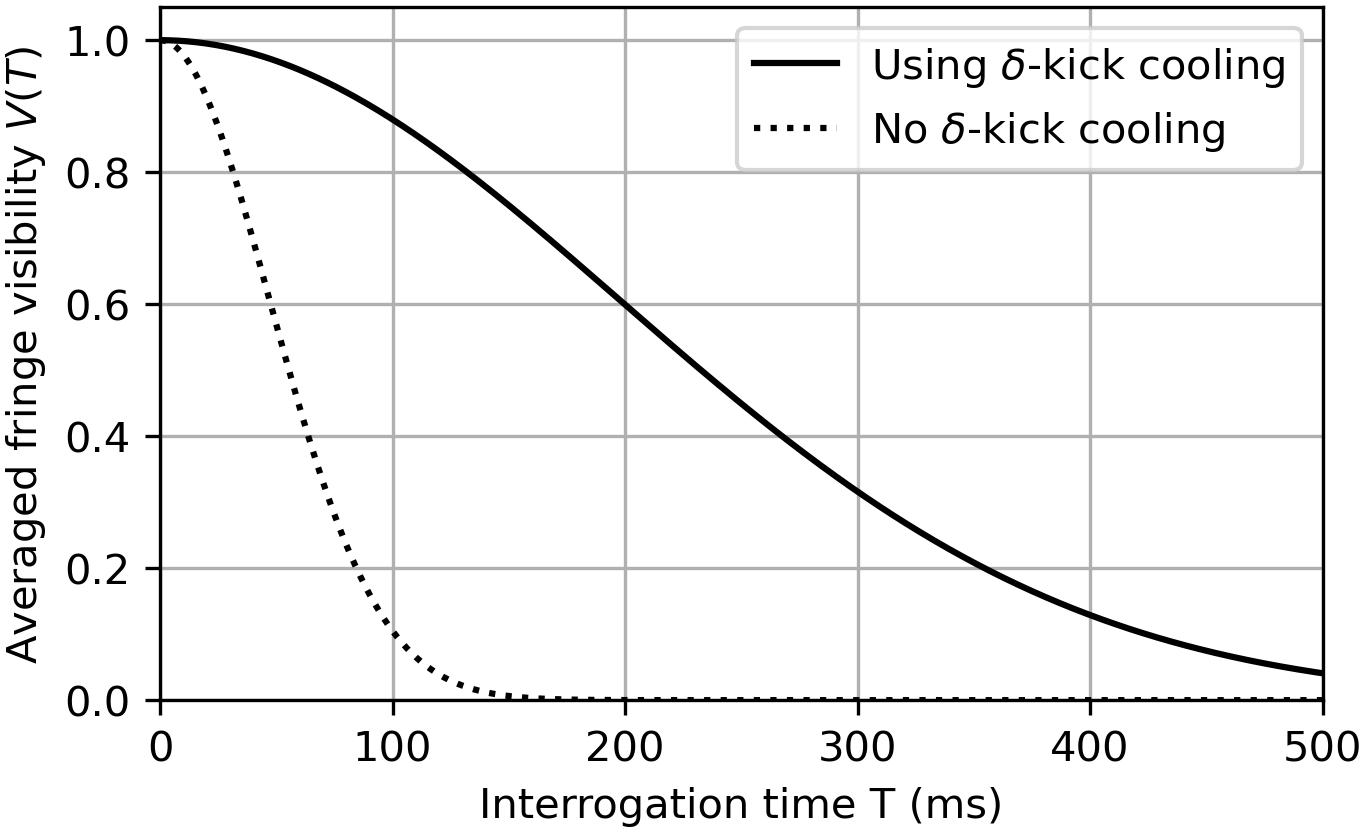}
\caption{\label{fig:PhaseDiffusion} The effect of phase diffusion on the fringe visibility as a function of the total interrogation time $T$, for $N=1000$ atoms. The solid curve shows fringe visibility for an expanded $\delta$-kicked cooled atomic cloud, while the dashed line is calculated for the cloud before the expansion. Both calculations are based on Eq.~(\ref{eq:PhaseDiffCalc}) and assume constant size of the cloud during the interrogation time.}
\end{figure}

We calculated the accumulated phase $\xi$ from the density profile of the cloud $|\psi_\pm|^2 = n(\mathbf{r})$, which was measured experimentally. Since, in our case, the wave-packet size does not change significantly during the interrogation time due to the delta-kick cooling technique, we can assume constant size of the split clouds. The formula used to obtain fringe visibility $V(T)$ versus interrogation time $T$ in Fig.~\ref{fig:PhaseDiffusion} takes the following form:
\begin{equation}\label{eq:PhaseDiffCalc}
 \xi(T) = \frac{U_0 }{2 \hbar} T \int n(\mathbf{r})^2 \mathbf{dr}
\end{equation}
Figure~\ref{fig:PhaseDiffusion} shows two cases: fringe visibility for an expanded (delta-kick cooled) cloud with FWHM $\approx \SI{50}{\micro \metre}$, and for a cloud before the expansion with FWHM $\approx \SI{15}{\micro \metre}$.
In the first case, the phase diffusion effect is not significant for the interrogation times up to $T=\SI{80}{\milli \second}$ used in the experiment. However, fringe visibility for the denser cloud present before expansion decreases significantly faster. Even though the calculation for the small cloud is less accurate because the assumption of constant size is not valid, it is still a useful reference point, indicating that the delta-kick cooling technique is essential for obtaining high contrast interferometer fringes.

\bibliography{References.bib}

\end{document}